\documentclass[12pt]{article}
\usepackage{epsfig}

\def\GG{\gamma\gamma}
\def\GE{e\gamma}

\begin{document}
\newpage

\begin{flushright}
SLAC-PUB-10410\\
April 2004
\end{flushright}

\bigskip\bigskip

\begin{center}
{{\bf\Large HIGH ENERGY PHOTON--PHOTON\\[1.5ex] COLLISIONS AT A LINEAR
COLLIDER\footnote{Work supported by the Department of Energy under
constract number DE--AC03--76SF00515.}}}

\vfill
Stanley J. Brodsky \\
Stanford Linear Accelerator Center, Stanford University \\
Stanford, California 94309 \\
e-mail: sjbth@slac.stanford.edu

\bigskip

\end{center}

\vfill

\begin{center}
Invited talk, presented at the\\
 5th International Workshop On
Electron-Electron Interactions At TeV Energies\\
Santa Cruz, California\\
 12--14 December 2003

\end{center}

\vfill

\newpage

\begin{abstract}
High intensity back-scattered laser beams will allow the efficient conversion of a
substantial fraction of the incident lepton energy into high energy photons, thus
significantly extending the physics capabilities of an $e^- e^\pm$ linear
collider. The annihilation of two photons produces $C=+$ final states in virtually
all angular momentum states. An important physics measurement is the measurement
of the Higgs coupling to two photons.  The annihilation of polarized photons into
the Higgs boson determines its fundamental $H^0 \to \gamma \gamma$ coupling as
well as determining its parity. Other novel two-photon processes include the
two-photon production of charged pairs $\tau^+ \tau^-,$ $W^+ W^-$, $t \bar t,$ and
supersymmetric squark and slepton pairs. The one-loop box diagram leads to the
production of pairs of neutral particles such as $\gamma \gamma \to Z^0 Z^0 ,
\gamma Z^0,$ and $\gamma \gamma.$ At the next order one can study Higgstrahlung
processes, such as $\gamma \gamma \to W^+ W^- H.$ Since each photon can be
resolved into a $W^+ W^-$ pair, high energy photon-photon collisions can also
provide a remarkably background-free laboratory for studying possibly anomalous $W
W$ collisions and annihilation. In the case of QCD, each photon can materialize as
a quark anti-quark pair which interact via multiple gluon exchange. The
diffractive channels in photon-photon collisions allow a novel look at the QCD
pomeron and odderon. The $C=-$ odderon exchange contribution can be identified by
looking at the heavy quark asymmetry. In the case of $e \gamma \to e^\prime$
collisions, one can measure the photon structure functions and its various
components. Exclusive hadron production processes in photon-photon collisions
provide important tests of QCD at the amplitude level, particularly as measures of
hadron distribution amplitudes which are also important for the analysis of
exclusive semi-leptonic and two-body hadronic $B$-decays.
\end{abstract}

\bigskip

\section{Introduction}
One of the important areas of investigation at the proposed  high
energy electron-positron linear collider will be the study of
photon-photon collisions. Since photons couple directly to all
fundamental fields carrying the electromagnetic current---%
leptons, quarks, $W's,$ supersymmetric particles, etc.---%
high energy $\gamma \gamma $ collisions will provide a comprehensive laboratory
for exploring virtually every aspect of the Standard Model and its
extensions~\cite{Velasco:2002vg,Asner:2003hz,DeRoeck:2003gv,Krawczyk:2003yz,Urner:2003gi,Brodsky:1994nf,Brodsky:2002zk}.
Effective photon beams from virtual bremsstrahlung provide access to low energy
quasi-real $\gamma \gamma$
collisions~\cite{Brodsky:1971ud,Budnev:de,Terazawa:tb,Defrise:1980gf}. A large
number of studies have been performed at the 4 detectors at LEP, at CESR, BaBar,
Belle, VEPP-4, VEPP-2, Adone,  and other $e^+ e^-$ storage rings. The QED
processes $e^+ e^- \to e^+ e^- \mu^+ \mu^-$ and $e^+ e^-  \to e^+ e^- \tau^+
\tau^-$ have been studied at the L3 detector at LEP at  161 GeV $< \sqrt{s} <$ 209
GeV.  The muon pair invariant mass was measured in the range 3 GeV $<
W_{\gamma\gamma}< $ 40 GeV. Good agreement was found  with ${\cal O}(\alpha^4)$
QED expectations. In addition, limits on the anomalous magnetic and electric
dipole moments of the tau lepton were obtained~\cite{Achard:2004jj}.

\begin{figure}[htbp]
 \begin{center}
 \mbox{\epsfig{file=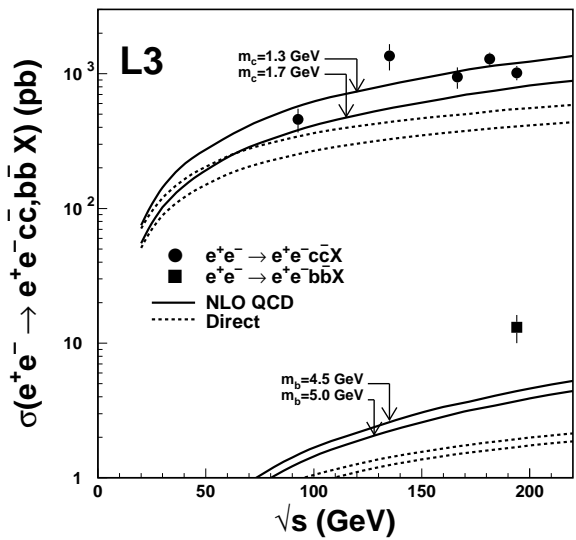,width=3.25in}} 
    \caption[*]{The open charm and beauty production cross section
      in two-photon collisions. The
      L3 data from both electron and muon events
      are combined. The statistical and systematic uncertainties
      are added in quadrature. The dashed line corresponds
      to the direct process contribution and the solid
      line represents the next-to-leading-order QCD prediction for the sum of
      the direct and resolved processes.}
      \end{center}
    \label{fig:sigma_ccbb}
\end{figure}

It should be noted  that significant deviations from QCD predictions have been
reported by the L3 and OPAL collaborations at LEP for the cross sections of
inclusive single-particle, jet, and beauty
production~\cite{Acciarri:2000kd,L3-p0k0,L3-cha,OPAL-chak0} in $\gamma \gamma$
collisions. A recent review of these measurements has been given by
Braccini~\cite{Braccini:2003rm}. For example, the production of $c$ and $b$ quarks
in $\gamma-\gamma$ collisions has been studied by L3 at $W_{\gamma \gamma} =\sqrt
s_{\gamma \gamma}$ from $189$ GeV to $202$ GeV. The measured cross sections are in
excess of the next-to-leading order perturbative QCD predictions by a factor of
3~\cite{Acciarri:2000kd}. See Fig. \ref{fig:sigma_ccbb}.

\begin{figure}[t]
\begin{center}
\psfig{file=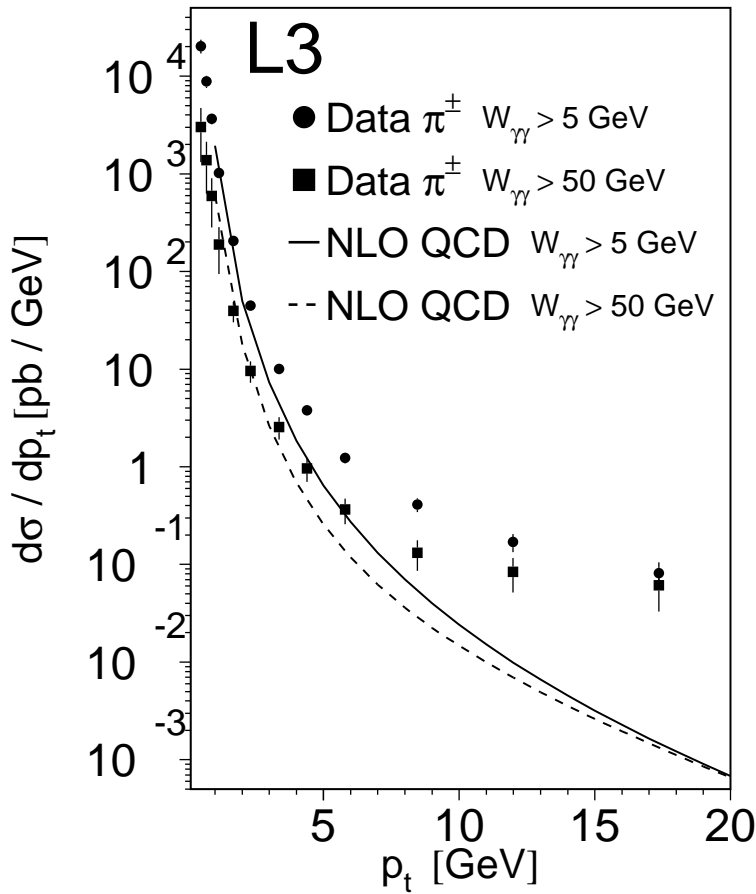,width=.65\textwidth}
\end{center}
\caption{L3 measurement of inclusive charged
hadron production in two-photon collisions compared with QCD predictions.}
\label{fig02} 
\end{figure}

The cross section for $\pi^\pm$, $\pi^0$ and ${\rm K}^0_s$
inclusive production in the reaction $\gamma\gamma\rightarrow$
hadrons has been measured as a function of the transverse momentum
and the pseudorapidity by L3~\cite{L3-p0k0,L3-cha} and
OPAL~\cite{OPAL-chak0}. Agreement with respect to QCD predictions
is found for all particles at $p_t < 4$ GeV; however the data are
significantly higher than the QCD predictions for $\pi^\pm$ and
$\pi^0$ at high $p_t$, as shown in Fig.~\ref{fig02}.

The advent of back-scattered laser beams for $e^\pm e^-$ colliders
will allow the efficient conversion of a substantial fraction of
the incident lepton energy into high energy
photons~\cite{Ginzburg:1981ik,Telnov:1998vs}.  When a polarized
laser beam Compton-scatters on a polarized electron beam, each
electron is effectively converted into a polarized photon with a
high fraction of its energy. The technology for the required
high-powered lasers is well along in
development~\cite{Asner:2001vu}. The effective luminosity and
energy of photon-photon collisions from back-scattered laser beams
is expected to be comparable to that of the primary
electron-positron collisions.  Polarized electron-photon
collisions are also an important by-product of this
program~\footnote{An analogous QCD process---gluon-quark Compton
scattering---could have interesting consequences in heavy ion
collisions, where the repeated Compton backscattering of gluons on
colliding quarks can produce a ``gluon avalanche", thus providing
a dynamical mechanism for initiating a quark gluon
plasma~\cite{Brodsky:2003mm}.}

\begin{figure}[htbp]
\begin{center}
\psfig{file=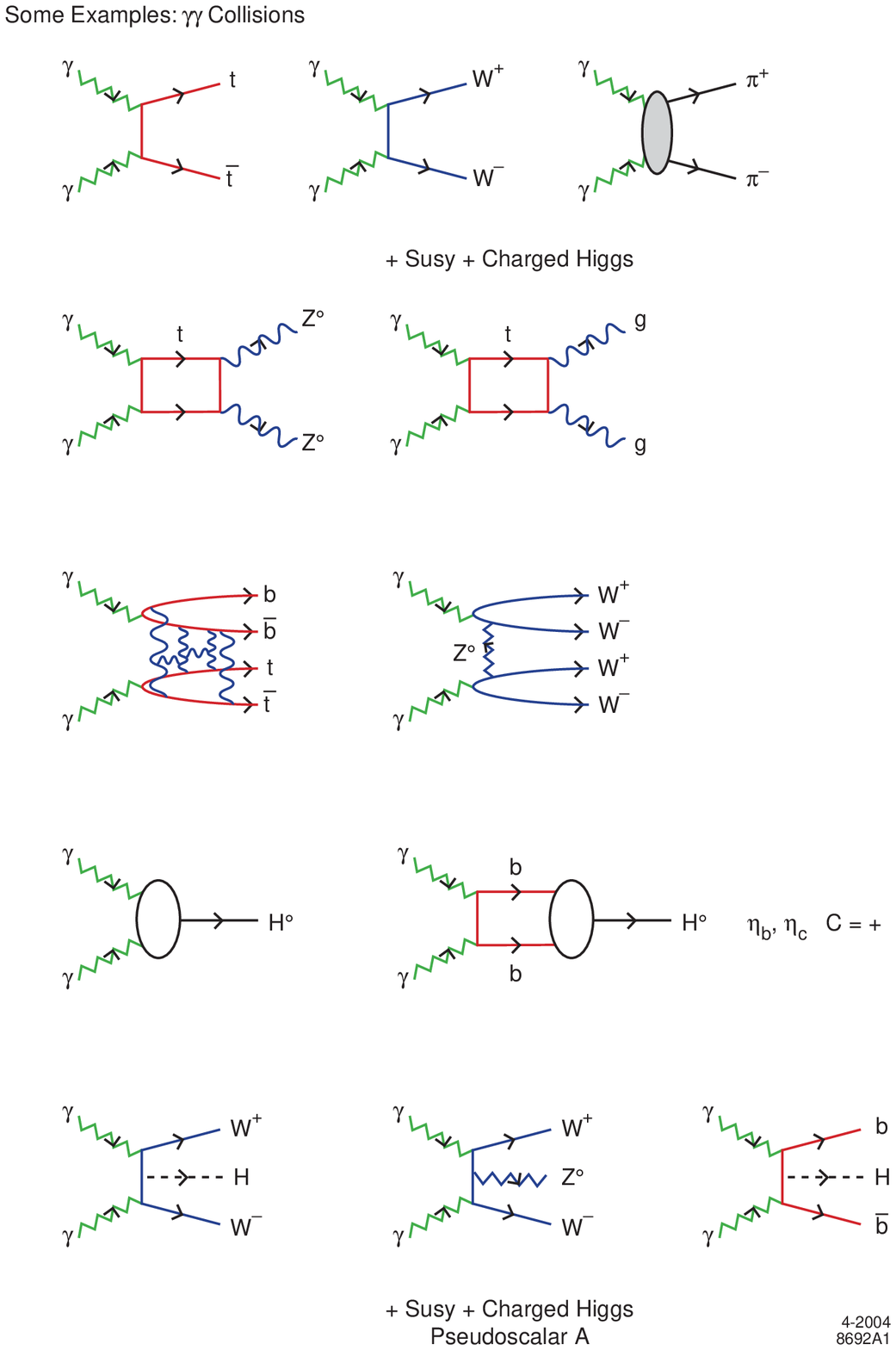,height=6.4in,width=4.5in}
 \end{center}
\caption[*]{Representative $\gamma \gamma$  processes accessible at
 a high energy photon-photon collider.}\label{fig:Feynmans}
\end{figure}

The high energy luminosity, and polarization of back-scattered laser beams thus
has the potential to make photon-photon collisions a key component of the physics
program of the next linear collider~\cite{Velasco:2002vg,Asner:2003hz}. This
capability will allow detailed studies of a large array of high energy $\gamma
\gamma$ and $\gamma e$ collision processes, including polarized beams. The physics
program includes tests of electroweak theory in photon-photon annihilation such as
$\gamma \gamma \to W^+ W^-$, $\gamma \gamma \to $ neutral and charged Higgs
bosons, and higher-order loop processes, such as $\gamma \gamma \to \gamma \gamma,
Z \gamma, H^0 Z^0$ and $Z.$ Since each photon can be resolved into a $W^+ W^-$
pair, high energy photon-photon collisions can also provide a remarkably
background-free laboratory for studying $W W$ collisions and annihilation.  There
are also important high energy $\gamma \gamma$ and $e \gamma$ tests of quantum
chromodynamics, including the production of two gluon jets in photon-photon
collisions, deeply virtual Compton scattering on a photon target, and
leading-twist single-spin asymmetries for a photon polarized normal to a
production plane. Exclusive hadron production processes in photon-photon
collisions provide important tests of QCD at the amplitude level, particularly as
measures of hadron distribution amplitudes which are also important for the
analysis of exclusive semi-leptonic and two-body hadronic $B$-decays.  Some of
these processes are illustrated in Fig.~\ref{fig:Feynmans}.

\begin{figure}[htb]
\begin{center}
\psfig{file=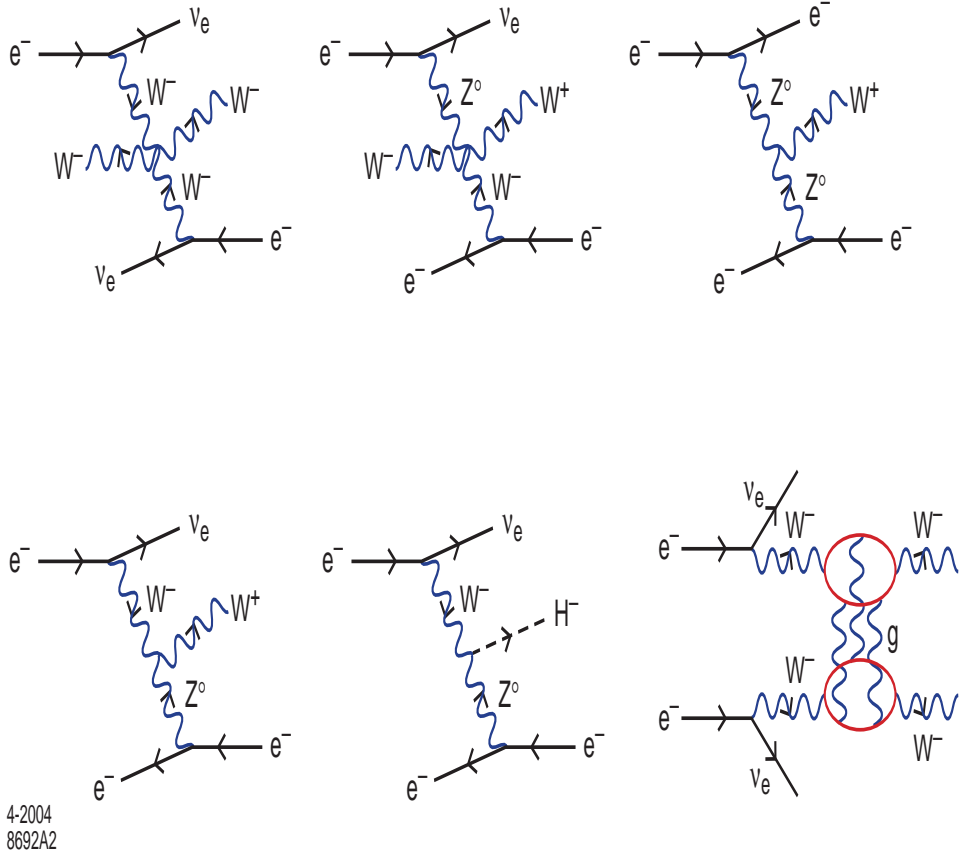,height=4.6in,width=5in}
 \end{center}
\caption[*]{Representative $e^- e^-$  processes.}\label{fig:eeprocesses}
\end{figure}

The photon-photon collider is also compatible with $e^- e^-$ collisions which are
interesting in their own right~\cite{Heusch:yb}. Some of the interesting processes
are the production of single and double-charged Higgs bosons via $e^- e^- \to Z^0
e^- W^- \nu \to H^- e^- \nu$ and
 $e^- e^- \to W^- \nu W^- \nu \to \nu \nu H^{--};$
 studies of $Z^0 Z^ 0 \to H^0$ fusion in
 $e^- e^- \to Z^0 e^- Z^0 e^- \to H^0 e^- e^-;$
 measurements of elastic vector boson scattering in
 $e^- e^- \to W^- \nu W^- \nu \to \nu \nu W^- W^-;$
including tests of the 3- and 4-point  $Z^0 Z^0 \to Z^0,$ $W^- Z^0 \to W^-,$ and
$W^- W^- \to W^- W^-$ couplings via
 $e^- e^- \to Z^0 e^- Z^0 e^- \to Z^0 e^- e^-$ and
  $e^- e^- \to Z^0 e^- W^- \nu \to W^- e^- \nu;$
 studies of the $W$ QCD structure function in
 $e^- e^- \to e^- \gamma^* W^- \nu  \to q \bar q \nu;$
 as well as measurements of high momentum transfer $e^- e^-$ elastic scattering.
 Some of these processes are illustrated in Fig.~\ref{fig:eeprocesses}.

Photon-photon collisions can be classified as follows: (A) The photons can
annihilate into a charged pair such as $\gamma \gamma \to W^+ W^-, q \bar q$,
lepton pairs or charged Higgs; (B) the photons can produce neutral pairs via loop
diagrams such as $\gamma \gamma \to Z^0 Z^0, \gamma Z^0$  and $\gamma \gamma \to g
g$ ; or (C) the photons can each couple to separate charged pairs which scatter by
a gauge particle exchange: $\gamma \gamma \to q_1 \bar q_1 q_2 \bar q_2;$ (D) the
photons can fuse to produce a single $C=+$ resonance such as a neutral Higgs, an
$\eta_b,$ or $\chi_b$ higher orbital state. Exclusive hadronic final states such
as meson or baryon pairs can be formed.  In each case, a state of even charge
conjugation $C$ is produced in a general partial wave. A recent survey of the
physics potential of $e e$ $e \gamma$ and $\gamma \gamma $ colliders has been
given by De Roeck~\cite{DeRoeck:2003gv}.  A detailed study of Higgs production in
the Standard Model and the Minimal Supersymmetric Standard model (MSSM) has been
summarized by Krawczyk~\cite{Krawczyk:2003yz} and Asner~\cite{Asner:yf}. Probes of
the physics of alternative models, such as the ``little Higgs"
model~\cite{Arkani-Hamed:2002qy} is discussed by Asner et al.~\cite{Asner:2003hz}.
A review of recent experimental results in two-photon interactions is given by
Urner~\cite{Urner:2003gi}.

Table I summarizes some of the important processes accessible at a
photon collider, as itemized by Boos {\it et
al.}~\cite{Boos:2000ki}  One can add processes such as $e\gamma
\rightarrow e^*$, leptoquark production, strong $ W W$ scattering,
and $e\gamma \to eH$.

\begin{table}[htb] 
\caption{Update of the Gold--plated processes at photon colliders.}
{\begin{tabular}{@{}l  c@{} }
\hline
$\quad$ {\bf Reaction} & {\bf Remarks} \\
\hline\hline
$\GG\to  H,h\to b\bar b$ & SM/MSSM\  Higgs,
 $M_{H,h}<160$~GeV \\
$\GG\to H\to WW(^*)$    & SM\ Higgs,
140$< M_{H}<190$~GeV \\
$\GG\to H \to ZZ(^*)$      & SM\
Higgs,  180$< M_{H}<350$~GeV \\
$\GG\to H \to \gamma\gamma$      & SM\
Higgs,  120$< M_{H}<160$~GeV \\
$\GG\to H \to t\overline{t}$      & SM\
Higgs,  $ M_{H}>350$ GeV \\
\hline
$\GG \to H,A\to b\bar b$  &
 MSSM\ heavy Higgs,  intermediate. $\tan\beta$\\
$\GG\to \tilde{f}\bar{\tilde{f}},\
\tilde{\chi}^+_i\tilde{\chi}^-_i$ & large cross sections \\
$\GG\to \tilde{g}\tilde{g}$ & measurable cross sections\\
$\GG\to  H^+H^-$ & large cross sections \\
$\GG\to S[\tilde{t}\bar{\tilde{t}}]$ &
$\tilde{t}\bar{\tilde{t}}$ stoponium  \\
$\GE \to \tilde{e}^- \tilde{\chi}_1^0$ &
 $M_{\tilde{e}^-} < 0.9 \times 2E_0 - M_{\tilde{\chi}_1^0}$  \\
\hline
$\gamma\gamma \to \gamma\gamma$ &   non-commutative theories \\
$e\gamma \to eG $ &   extra dimensions\\
$\gamma\gamma  \to \phi$    &   Radions \\
$e\gamma \to \tilde{e}\tilde{G} $ & superlight gravitions\\
\hline
$\GG\to W^+W^-$ & anom. $W$ inter., extra dimensions \\
$\GE\to W^-\nu_{e}$ & anom.$W$ couplings \\
$\GG\rightarrow 4W/(Z)$& $WW$ scatt.,
quartic anom.~$W$,$Z$\\
\hline
$\GG\rightarrow t\bar{t}$ & anomalous top quark interactions \\
$\GE\rightarrow \bar t b \nu_e$ & anomalous $W tb$ coupling \\
\hline
$\GG\rightarrow$ hadrons & total $\GG $ cross section \\
$\GE\rightarrow e^- X$, $\nu_{e}X$ & NC and CC structure functions \\
$\gamma g\rightarrow q\bar{q},\ c\bar{c}$ & gluon in the photon \\
$\GG\to J/\psi\, J/\psi $ & QCD Pomeron \\
\hline
\end{tabular}}
\end{table}

\begin{figure}[htb]
\begin{center}
\psfig{file=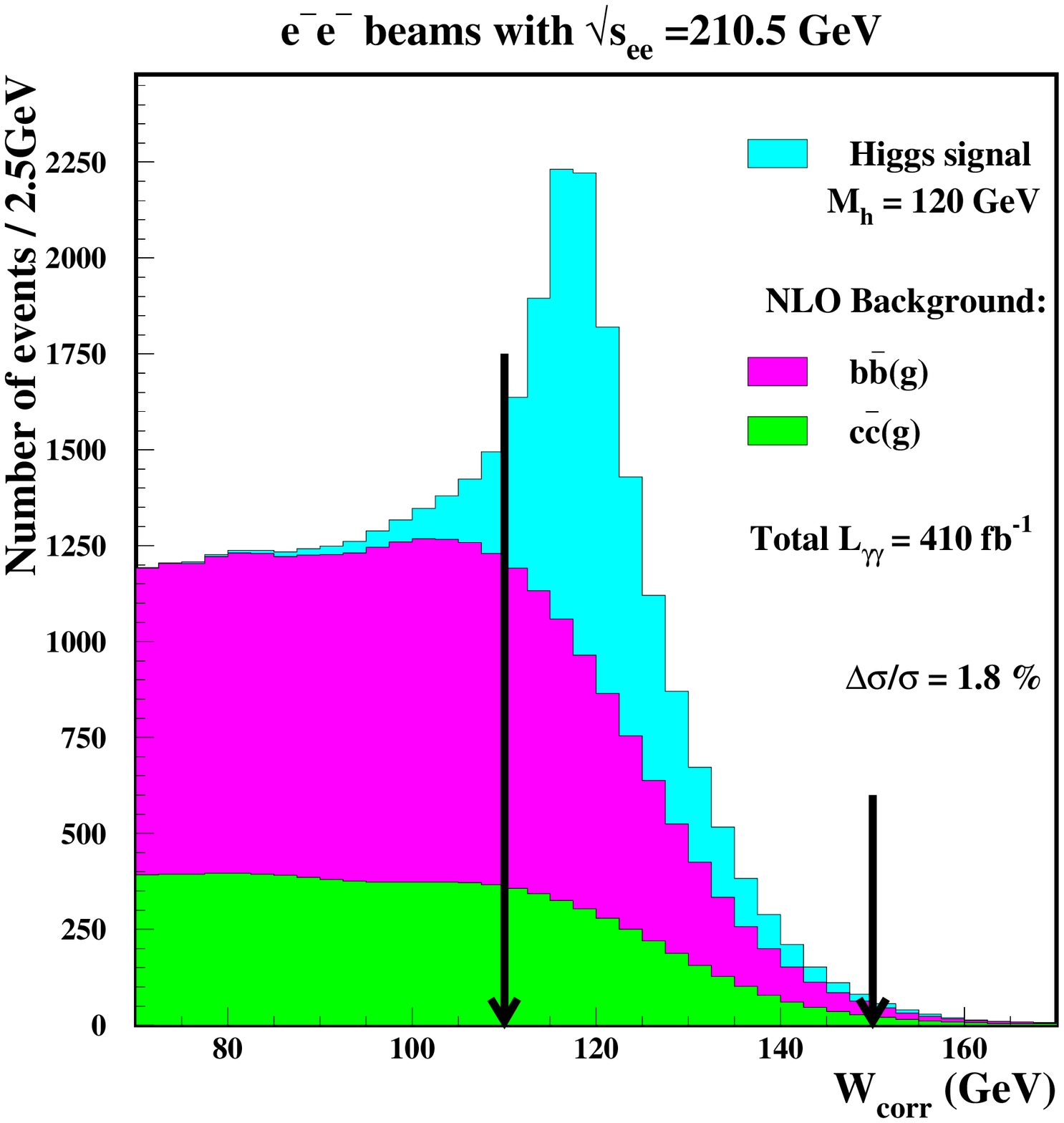,bbllx=0pt,bblly=0pt,bburx=540pt,bbury=570pt,height=9.5cm}
 \end{center}
 \caption[*]{Simulation of Higgs events $H^0\to b \bar b$ and the related
background events at a photon-photon collider, assuming $M_H= 120$ GeV.  The rate
for photon-photon collisions will determine the quantity $\Gamma(H \to \gamma
\gamma ) \times {\rm BR}(H \to b \bar b)$.  The arrows show the optimized mass
window for the partial width measurement. See
references~\cite{DeRoeck:2003gv,piotr,rosca}.} \label{fig:higgs1}
\end{figure}

\begin{figure}[htbp]
\begin{center}
\psfig{file=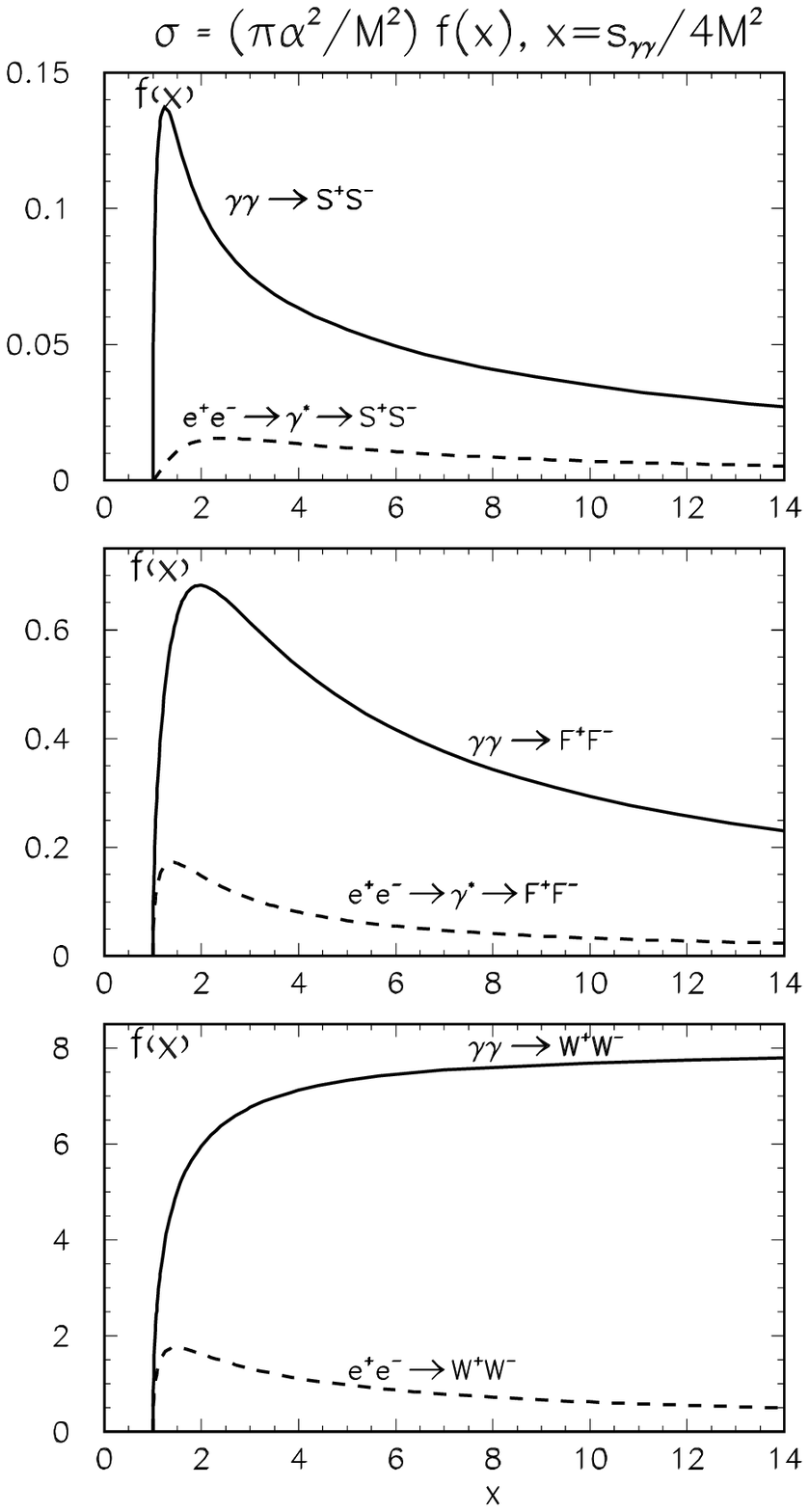,height=15cm, width=10cm} 
\end{center}
\caption[*]{Comparison between cross sections for charged pair production in
unpolarized $e^+ e^-$ and $\gamma \gamma$ collisions. S (scalars), F (fermions), W
($W$ bosons); $\sqrt{s}$ is the invariant mass (c.m.s. energy of colliding beams).
The contribution of the $Z^0$ boson to the production of S and F in $e^+ e^-$
collisions was not included. From Boos {\em et
al.}~\cite{Boos:2000ki}\label{charged}}
\end{figure}

A unique advantage of a photon-photon collider is its potential to produce and
determine the properties of fundamental $C=+$ resonances such as the Higgs boson.
A simulation of events for $\gamma \gamma \to H^0 \to b \bar b$ is shown in
Fig.~\ref{fig:higgs1}. One can also use the transverse polarization of the
colliding photons to distinguish the parity of the resonance: the coupling for a
scalar resonance is $\epsilon_1\cdot\epsilon_2$ versus $\epsilon_1\times
k_1\cdot\epsilon_2$ for the pseudoscalar.  More generally, one can use polarized
photon-photon scattering to study CP violation in the fundamental Higgs to
two-photon couplings~\cite{Grzadkowski:1992sa,Cheung:bn,Asner:2001ia}.  In the
case of electron-photon collisions, one can use the transverse momentum fall-off
of the recoil electron in $ e \gamma\rightarrow e H^0$ to measure the fall-off of
the $\gamma \to $ Higgs transition form factor and thus check the mass scale of
the internal massive quark and $W$ loops coupling to the Higgs~\cite{tang}. The
cross sections for pairs of scalars, fermions or vectors particles are all
significantly larger (by about one order of magnitude) in $\gamma \gamma$
collisions than in $e^+e^-$ collisions, as demonstrated in Fig.~\ref{charged}.

Unlike the $e^+ e^-$ annihilation cross section, which falls at
least as fast as $1/s,$ many of the $\gamma \gamma$ cross sections
increase with energy. The energy dependence of a cross section
follows from the spin of the exchanged quanta.  Using Regge
analysis, a two-body cross section ${d\sigma\over dt} \propto s^{2
\alpha_R(t)-2}\beta(t)$ at fixed $t$ where $\alpha_R$ is the spin
of the exchanged particle or effective trajectory.  For example,
the $\gamma \gamma \to W^+ W^-$ differential cross section is
constant at high energies since the spin of the exchanged $W$ is
$\alpha_R = j = 1.$ In fact, after integration over phase space,
the cross section for pairs of vector bosons in photon-photon
collisions increases logarithmically with energy.  This is in
contrast to $\sigma(e^+ e^- \to W^+ W^-)$ which produces a single
$W^\pm$ pair in one partial wave and falls as $1/s.$ It is also
interesting that a dominant two-jet high $p_T$ reaction in
photon-photon collisions at high energies $s \gg p^2_T$ is $\gamma
\gamma \to g g$ which proceeds via two quark loops coupling via
gluon exchange in the $t$ channel~\cite{HwangSjb}.

\section{Standard Model Tests}

\begin{figure}[htbp]
\begin{center}
\epsfig{file=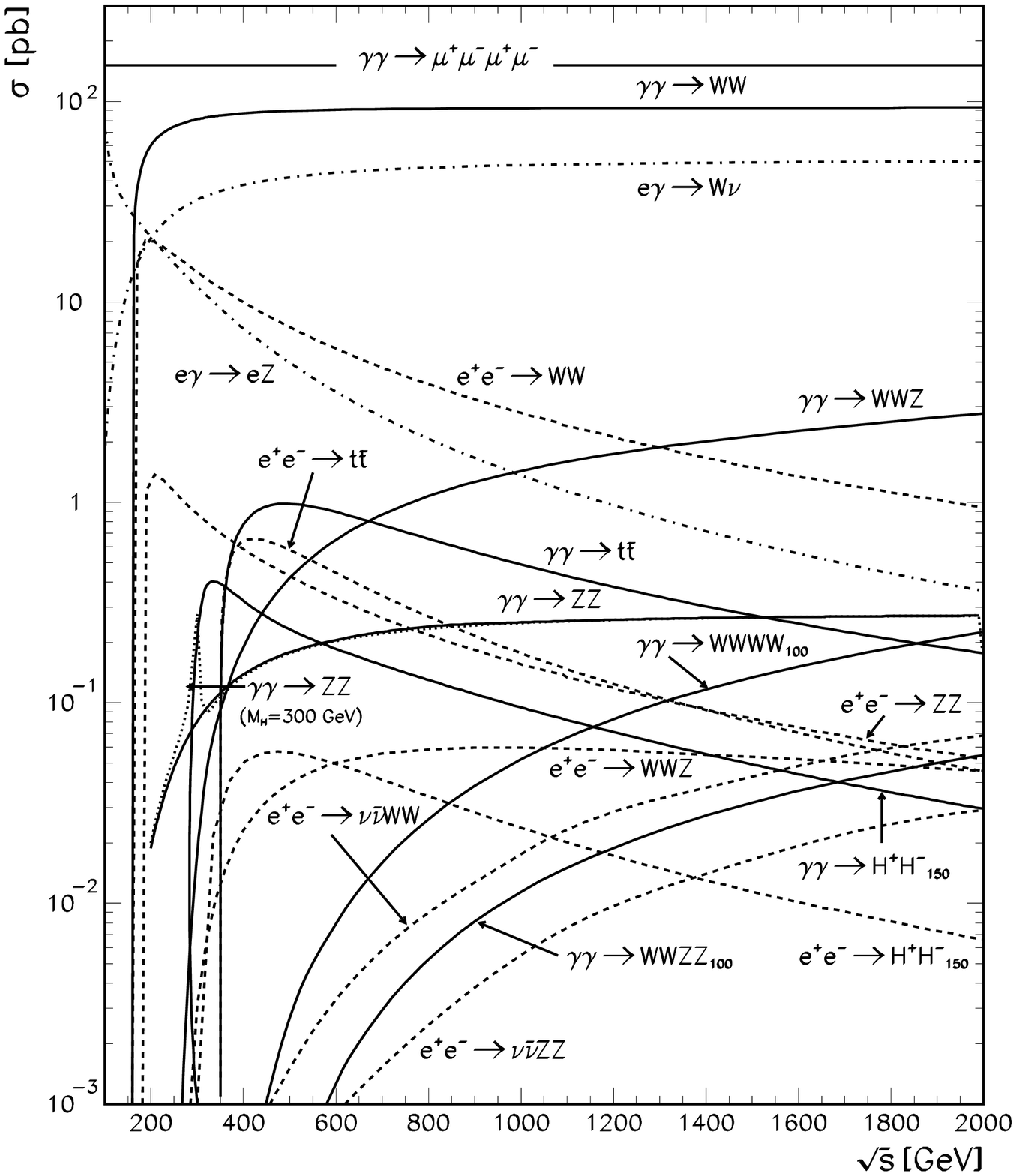,width=0.8\textwidth,height=0.8\textwidth}
\end{center}
\caption[*]{ Typical (unpolarized) cross sections in
$\gamma\gamma$, $\gamma e$ and $e^+e^-$ collisions. Solid,
dash-dotted and dashed curves correspond to $\gamma\gamma$,
$\gamma e$ and $e^+e^-$ modes respectively.  Unless indicated
otherwise the neutral Higgs mass was taken to be 100~GeV.  For
charged Higgs pair production, $M_{H^\pm}=150$~GeV was assumed.
From Boos {\em et al.}~\cite{Boos:2000ki} \label{fig:cs}}
\end{figure}

Since each photon can be resolved into a $W^+ W^-$ pair, high
energy photon-photon collisions produce equivalent effective
$W^\pm$ beams, thus providing a remarkably background-free
laboratory for studying $W W$ interactions and testing for any
anomalous magnetic and quadruple couplings.  The interacting
vector bosons can scatter pair-wise or annihilate; {\em e.g.},
they can annihilate into a Standard Model Higgs boson or a pair of
top quarks. There is thus a large array of tests of electroweak
theory possible in photon-photon collisions.  The splitting
function for $\gamma\to W^+ W^-$ can be relatively flat for some
$W$ helicities, so that one has a high probability for the $W$'s
to scatter with a high fraction of the energy of the photon. One
can thus study tree graphs contributions derived from photon, Z,
or Higgs exchange in the $t$-channel, and in the case of identical
$W$'s, the additional u-channel amplitudes.  In the case of
oppositely-charged $W$'s, $s$-channel annihilation processes such
as $W^+ W^- \to t\bar t$ contribute.  The largest cross sections
will arise if the $W$'s obey a strongly coupled theory; in this
case the longitudinal $W$'s scattering amplitude saturates
unitarity and the corresponding $\gamma\gamma \rightarrow WWWW$
cross section will be maximal. The cross sections of many Standard
Model processes are illustrated in Fig.~\ref{fig:cs}. Reviews of
this physics are given in the
references.~\cite{Boos:2000ki,Asner:2001vh,Brodsky:1993xp,Chanowitz:1994aq,Gunion:1992ce}

One of the most important applications of two-photon physics is the direct
production of $W^\pm$ pairs.  By using polarized back-scattered laser beams, one
can in principle study $\gamma \gamma \to W^+ W^-$ production as a function of
initial photon helicities as well as resolve the $W$ helicities through their
decays.  The study of $\gamma \gamma \to W^+W^-$ is complimentary to the
corresponding $e^+ e^- \rightarrow W^+W^-$ channel, but it also can check for the
presence of anomalous four-point $\gamma \gamma \to WW$ interactions not already
constrained by electromagnetic gauge invariance, such as the effects due to
$W^\ast$ exchange.

A main focus of the pair production measurements are the values of the $W$
magnetic moment $ \mu_W = {e\over 2m_W}\ (1-\kappa-\lambda)$ and quadruple
moment
$Q_W= - {e\over M^2_W}\ (\kappa-\lambda).$ The Standard Model predicts
$\kappa=1$
and $\lambda=0,$ up to radiative corrections analogous to the Schwinger
corrections to the electron anomalous moment.  The anomalous moments are thus
defined as $\mu_A = \mu_W-{e\over M_W}$ and $Q_A = Q_W + {e\over M^2_W}.$

The fact that $\mu_A$ and $Q_A$ are close to zero is actually a
general property of any  spin-one system if its size is small
compared to its Compton scale.  For example, consider the
Drell-Hearn-Gerasimov sum
rule~\cite{Drell:1966jv,Gerasimov:1965et} for the $W$ magnetic
moment: $ \mu^2_A = \left(\mu-{e\over M}\right)^2 =
{1\over\pi}\int^\infty _{\nu_{th}} {d\nu\over\nu}\,
[\sigma_P(\nu)-\sigma_A(\nu)]. $ Here $\sigma_{P(A)}$ is the total
photoabsorption cross section for photons on a $W$ with (anti-)
parallel helicities. As the radius of the $W$ becomes small, or
its threshold energy for inelastic excitation becomes large, the
DHG integral and hence $\mu^2_A$ vanishes. Hiller and I have
shown~\cite{Brodsky:1992px} that this argument can be generalized
to the spin-one anomalous quadruple moment as well, by considering
one of the unsubtracted dispersion relations for near-forward
$\gamma$ spin-one Compton scattering~\cite{Tung:kn}:
\begin{eqnarray}
&& \mu_A^2 + {2t\over M^2_W}\  \left(\mu_A+{M_2\over W}\
Q_A\right)^2 = \nonumber \\
&& {1\over 4\pi} \int^\infty_{\nu_{th}} {d\nu^2\over (\nu-t/4)^3}\
Im\, (f_P(s,t)-f_A(s,t))\ .
\end{eqnarray}
Here $\nu = (s-u)/4$.  One again sees that in the point-like or
high threshold energy limit, both
 $\mu_A \rightarrow 0,$ and
$Q_A\rightarrow 0.$ This result applies to any spin-one system,
even to the deuteron or the $\rho.$ The essential assumption is
the existence of the unsubtracted dispersion relations; {\em
i.e.}, that the anomalous moments are in principle computable
quantities.

In the case of the $W$, the finite size correction is expected to
be order $m^2/\Lambda^2$, since the underlying composite theory
should be chiral to keep the $W$ mass finite as the composite
scale $\Lambda$ becomes large~\cite{Brodsky:1980zm}. Thus the fact
that a spin-one system has nearly  canonical values for its
moments signals that it has a small internal size; however, it
does not necessarily imply that it is a gauge field.

Yehudai~\cite{Yehudai:1991az} has made extensive studies of the
effect of anomalous moments on different helicity amplitude
contributing to $\gamma\gamma \rightarrow W^+W^-$ cross section.
The empirical sensitivity to anomalous couplings from
$\gamma\gamma$ reactions is comparable and complimentary to that
of $e^+ e^- \rightarrow W^+W^-.$  A comprehensive analysis of
fermion processes in photon-photon collisions is given by Layssac
and Renard~\cite{Layssac:2001ur}.

As emphasized by Jikia~ and Tkabladze~\cite{Jikia:1993pg}, pairs
of neutral gauge bosons can be produced in $\gamma\gamma$
reactions through one loop amplitudes in the Standard Model at a
rate which should be accessible to the NLC. Leptons, quarks, and
$W$ all contribute to the box graphs. The fermion and spin-one
exchange contributions to the $\gamma \gamma \to \gamma \gamma$
scattering amplitude have the characteristic behavior ${\cal M}
\sim s^0 f(t)$ and ${\cal M} \sim i\,s f(t)$ respectively.  The
latter is the dominant contribution at high energies, so one can
use the optical theorem to relate the forward imaginary part of
the scattering amplitude to the total $\gamma \gamma \to W^+ W^-$
cross section.  The resulting cross section $\sigma(\gamma\gamma
\rightarrow \gamma\gamma)$ is of order 20 fb at $\sqrt s_{\gamma
\gamma}$ , corresponding to 200 events/year at an NLC with
luminosity 10 fb$^{-1}$~\cite{Boos:2000ki}. The corresponding
$\gamma \gamma \to H^0 Z^0$ process has been analyzed by Gounaris,
Porfyriadis and Renard~\cite{Gounaris:2001rk}.

\begin{figure}[htb]
\begin{center}
{ \epsfig{file=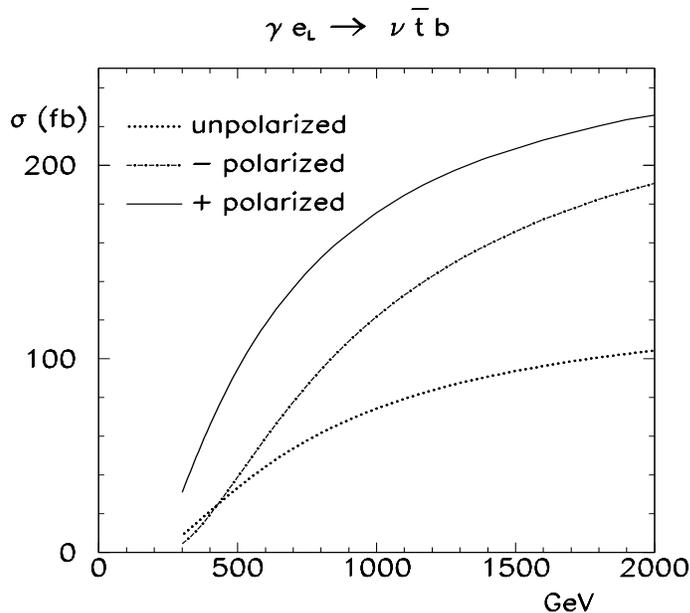,width=10cm,height=9cm}}
\end{center}
\caption[*]{Single top quark production cross section in $\gamma
e$ collisions. From Boos {\em et al.}~\cite{Boos:2001sj}
\label{fig:f9}}
\end{figure}

A single top quark can be produced in electron-photon collisions
at an NLC through the process $e^-\gamma \rightarrow W^-
t\nu$~\cite{Jikia:1991hc}. See
Fig.~\ref{fig:f9}.~\cite{Boos:2001sj} This process can be
identified through the $t \to W^+ b$ decay with $W\ \to \ell \bar
\nu.$ The rate is strongly polarization dependent and is sensitive
to the structure of the $V_{tb}$ matrix element, possible fourth
generation quarks, and anomalous couplings. An interesting
background is the virtual $W$ process $e\gamma\rightarrow W^\ast
-\nu \to W^- H \nu,$ where the Higgs boson decays to $b\bar b$ and
$W^-\to \ell \bar \nu.$

\begin{figure}[htb]
\begin{center}
\leavevmode {\epsfxsize=3.truein \epsfbox{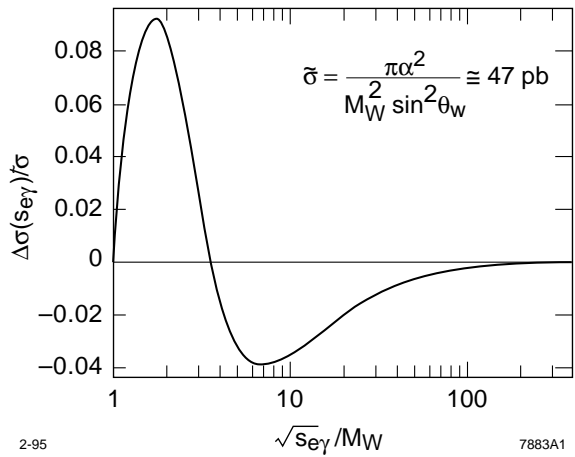}}
\end{center}
\caption[*]{The Born cross section difference $\Delta \sigma$ for
the Standard Model process $\gamma e \to W \nu$ for parallel minus
antiparallel electron/photon helicities as a function of $\log
\sqrt s_{e \gamma}/M_W$ The logarithmic integral of $\Delta
\sigma$ vanishes in the classical limit.~\cite{Brodsky:1995ga}
\label{figB}}
\end{figure}

Schmidt, Rizzo, and I~\cite{Brodsky:1995ga,Rizzo:1999xj} have
shown that one can use the sign change of the integrand of the DHG
sum rule to test the canonical couplings of the Standard Model and
to isolate the higher order radiative corrections. For example,
consider the reactions $\gamma \gamma \to q \overline q$, $\gamma
e \to W \nu$ and $\gamma e \to Z e$ which can be studied with
back-scattered laser beams.  In contrast to the time-like process
$e^+ e^- \to W^+ W^-$, the $\gamma \gamma$ and $\gamma e$
reactions are sensitive to the anomalous moments of the gauge
bosons at $q^2 = 0.$ The vanishing of the logarithmic integral of
$\Delta \sigma$ in the Born approximation implies that there must
be a center-of-mass energy, $\sqrt s_0$, where the polarization
asymmetry $A=\Delta \sigma/ \sigma$ possesses a zero, {\em i.e.},
where $\Delta \sigma({\gamma e \to W \nu })$ reverses sign. The
cancellation of the positive and negative contributions
\cite{ginz} of $\Delta \sigma(\gamma e \to W \nu)$ to the DHG
integral is evident in Fig.~\ref{figB}. We find strong sensitivity
of the position of this zero or ``crossing point'' (which occurs
at $\sqrt s_{\gamma e} = 3.1583 \ldots  M_W \simeq 254$ GeV in the
SM) to modifications of the SM trilinear $\gamma W W$ coupling and
thus can lead to high precision constraints.  In addition to the
fact that only a limited range of energy is required, the
polarization asymmetry measurements have the advantage that many
of the systematic errors cancel in taking cross section ratios.
This technique can clearly be generalized to other higher order
tree-graph processes in the Standard Model and supersymmetric
gauge theory.  The position of the zero in the photoabsorption
asymmetry thus provides an additional weapon in the arsenal used
to probe anomalous trilinear gauge couplings.

\vspace{.5cm}

\section{Inclusive QCD Tests}

Because of the simplicity of its initial state, two-photon
collisions provide an important laboratory for testing coherent
and incoherent effects in quantum chromodynamics. In QCD events
where each photon is resolved~\cite{Brodsky:1978rp,Drees:1995ti}
in terms of its intermediate quark and gluon states,  $\gamma
\gamma$ collisions resemble point-like meson-meson collisions. One
can study detailed features of $\gamma \gamma \to t \bar t$ at
threshold and its final state evolution.  In the case of single or
double diffractive two-photon events, one can study fundamental
aspects of pomeron and odderon $t$-channel physics.  For example,
the energy asymmetry of charm quark versus anti-charm jets in
four-jet reactions can measure the interference of the pomeron and
odderon~\cite{Brodsky:1999mz}.

Consider, the cross section for producing two quark pairs. $\gamma
\gamma \to q_1 \bar q_1 q_2 \bar q_2$, which can be described as
the interaction of two color
dipoles~\cite{Brodsky:1997sd,Bartels:1996ke}.  If one of the heavy
quark pairs is a charm or bottom quark pair with a small color
dipole moment, then the multiple gluonic exchange graphs will
Reggeize into a hard pomeron, and one predict a strong energy
growth of the production cross section similar to that observed at
HERA in diffractive charm electroproduction~\cite{Lipatov:2000me}.
The QCD description of the hard QCD pomeron is given by the BFKL
analysis~\cite{Fadin:1998py}.  There has been progress in
stabilizing the BFKL predictions at next-to-leading-order by using
BLM scale fixing~\cite{Brodsky:1998kn}.  The analogous QCD physics
of the gluon rescattering in diffractive deep inelastic scattering
is discussed in the references~\cite{Brodsky:2001ue}.

\section{The Photon Structure Functions}

One can also utilize electron-photon collisions at a linear
collider to test the shape and growth of the photon structure
functions~\cite{Brodsky:1971vm,Walsh:xy,Krawczyk:2000nh,Drees:1995ti}
The back-scattered laser beam provides a high energy polarized
target photon, and the neutral current probe is obtained by
tagging the scattered electron at momentum transfer squared $Q^2.$
One can also reconstruct the charged current contributions where
the electron scatters into a neutrino from calorimetric
measurements of the recoiling system. It also should be possible
to identify the separate charm, bottom, top and $W$ contributions
to the photon structure functions.

The photon structure functions receive hadron-like contributions
from the photon's resolved Fock components as well as its direct
component derived from the $\gamma^\ast \gamma \to q \bar q$~
time-like QCD Compton amplitude. Because of the direct
contributions, the photon structure functions obey an
inhomogeneous evolution equation. The result,  as first shown by
Witten~\cite{Witten:ju} is that the leading order QCD structure
functions of the photon have a unique scaling behavior:
$F_1(x,Q^2) = h(x)\, \ell n {Q^2\over \Lambda^2}$, $F_2(x,Q^2) =
f_2(x)$, $F_3(x,Q^2) = f^{\rm Box}_3(x)$.

The most characteristic behavior of the photon structure function
$F_2^\gamma(x,Q^2)$ in QCD is its continuous linear rise of with
$\log Q^2$ at fixed $x$.  As emphasized by Peterson, Walsh and
Zerwas~\cite{Peterson:1982tt}, the fact that this tree graph
behavior is preserved to all orders in perturbation theory is due
to the balance in QCD between the increase of the phase space for
gluon emission in the scattering processes versus the decreasing
strength of the gluon coupling due to asymptotic freedom. Although
the logarithmic rise of the Born approximation result is
preserved, the shape of $h(x)$ is modified by the QCD radiation.
If the running coupling constant were to freeze to a constant
value at large momentum transfer, the photon structure function
stops rising at high $Q^2$ due to the increased phase space for
gluon radiation. Thus probing the QCD photon structure functions
at the high momentum transfers available at the NLC will provide a
valuable test of asymptotic freedom.

\section{The
Photon Structure Function and Final-State Interactions}

Hoyer, Marchal, Peigne, and Sannino and I~\cite{Brodsky:2001ue}
have challenged the common view that structure functions measured
in deep inelastic lepton scattering are determined by the
probability of finding quarks and gluons in the target hadron.  We
show that this is not correct in gauge theory. Gluon exchange
between the fast, outgoing partons and target spectators, which is
usually assumed to be an irrelevant gauge artifact, affects the
leading twist structure functions in a profound way.  This
observation removes the apparent contradiction between the
projectile (eikonal) and target (parton model) views of
diffractive and small $x_{\rm Bjorken}$ phenomena.  The
diffractive scattering of the fast outgoing quarks on spectators
in the target in turn causes shadowing in the DIS cross section.
Thus the depletion of the nuclear structure functions is not
intrinsic to the wave function of the nucleus, but is a coherent
effect arising from the destructive interference of diffractive
channels induced by final-state interactions.  This is consistent
with the Glauber-Gribov interpretation of shadowing as a
rescattering effect.  Similar effects can be present in the photon
structure function; {\em i.e.}, the photon structure function will
be modified by rescattering of the struck quark with the photon's
spectator system.

\section{Single-Spin Asymmetries in
Photon-Photon Collisions}

Measurements from the HERMES and SMC collaborations show a
remarkably large azimuthal single-spin asymmetries $A_{UL}$ and
$A_{UT}$ of the proton in semi-inclusive pion leptoproduction
$\gamma^*(q) p \to \pi X.$  Dae Sung Hwang and Ivan
Schmidt and I~\cite{Brodsky:2002cx} have shown that final-state
interactions from gluon exchange between the outgoing quark and
the target spectator system lead to single-spin asymmetries in
deep inelastic lepton-proton scattering at leading twist in
perturbative QCD; {\em i.e.}, the rescattering corrections are not
power-law suppressed at large photon virtuality $Q^2$ at fixed
$x_{bj}$.  The existence of such single-spin asymmetries requires
a phase difference between two amplitudes coupling the proton
target with $J^z_p = \pm {1\over 2}$ to the same final-state, the
same amplitudes which are necessary to produce a nonzero proton
anomalous magnetic moment.  We have shown that the exchange of
gauge particles between the outgoing quark and the proton
spectators produces a Coulomb-like complex phase which depends on
the angular momentum $L^z$ of the proton's constituents and is
thus distinct for different proton spin amplitudes.  The
single-spin asymmetry which arises from such final-state
interactions does not factorize into a product of distribution
function and fragmentation function, and it is not related to the
transversity distribution $\delta q(x,Q)$ which correlates
transversely polarized quarks with the spin of the transversely
polarized target nucleon.  These effects highlight the unexpected
importance of final and initial state interactions in QCD
observables.

Final state interactions will also lead to new types of single
spin asymmetries in photon-photon collisions. For example, in
$\gamma^* \gamma \to \pi X$ and $\gamma^* \gamma \to {\rm jet} X$
we expect $T$-odd correlations of the type $\vec S_\gamma \cdot
\vec q \times \vec p$ where $\vec S_\gamma$ is the polarization of
the real photon, $\vec q$ is the beam direction of an incident
virtual photon, and $\vec p$ is the direction of a produced quark
or hadron.  The resulting asymmetry of the photon polarized normal
to the production plane will be leading twist.  As in the proton
target case, the single-spin asymmetry will be sensitive to
orbital angular momentum in the photon wavefunction and details of
the photon structure at the amplitude level.

\section{Single and Double Diffraction in Photon-Photon Collisions}

The high energies of a photon-photon collider will make the study
of double diffractive $\gamma \gamma \to V^0 V^0$ and
semi-inclusive single diffractive processes $\gamma \gamma \to V^0
X$ in the Regge regime $s \gg |t|$ interesting. Here $V^0 = \rho,
\omega\phi,J/\psi,\cdots$ If $|t|$ is taken larger than the QCD
confinement scale, then one has the potential for a detailed study
of fundamental Pomeron processes and its gluonic composition. As
in the case of large angle exclusive $\gamma \gamma$ processes,
the scattering amplitude is computed by convoluting the hard
scattering PQCD amplitude for $\gamma \gamma \to q \bar q q \bar
q$ with the vector meson distribution amplitudes.  The two gluon
exchange contribution dominates in the Regge
regime~\cite{Chernyak:xe}, giving a characteristic exclusive
process scaling law of order $ {d\sigma\over dt}\
(\gamma\gamma\rightarrow V^0V^0)\sim {\alpha^4_s(t) / t^6}.$
Ginzburg, Ivanov and Serbo~\cite{Ginzburg:gy} have emphasized that
the corresponding $\gamma\gamma\rightarrow $ pseudoscalar and
tensor meson channels can be used to isolate the Odderon exchange
contribution, contributions related at a fundamental level to
three gluon exchange.

In addition, the photon can diffractively dissociate into quark
pairs $\gamma e \to q \bar q e'$ by Coulomb scattering on the
incoming electron.  This measures the transverse derivative of the
photon wavefunction ${\partial \over \partial k_\perp} \psi_{q
\bar q}(x,k_\perp,\lambda_i).$ This is the analog of the E791
experiment at Fermilab~\cite{Ashery:1999nq} which resolved the
pion light-front wavefunction by diffractive dissociation $\pi A
\to q \bar q A'$ on a nuclear target.  The results of the
diffractive pion experiment are consistent with color
transparency, and the momentum partition of the jets conforms
closely with the shape of the asymptotic distribution amplitude,
$\phi^\mathrm{asympt}_\pi (x) = \sqrt 3 f_\pi x(1-x)$,
corresponding to the leading anomalous dimension solution
\cite{BrodskyLepage} to the perturbative QCD evolution equation.

\section{Other QCD Tests in Photon-Photon Collisions}

Two-photon annihilation $\gamma^*(q_1) \gamma^*(q_2) \to $ hadrons
for real and virtual photons can thus provide some of the most
detailed and incisive tests of QCD.  Among the processes of
special interest are:

\begin{enumerate}

\item the production of four jets such as $\gamma \gamma \to c
\bar c c \bar c$ can test Fermi-color statistics for charm quarks
by checking for the interference effects of like sign
quarks~\cite{BA}.

\item the total two-photon annihilation hadronic cross section
$\sigma(s,q^2_1,q^2_2),$ which is related to the light-by-light
hadronic contribution to the muon anomalous moment;

\item the formation of $C = +$ hadronic resonances, which can
reveal exotic states such as $q \bar q g$ hybrids and discriminate
gluonium formation \cite{Pennington:2000ai,Acciarri:2001ex}.  The
production of the $\eta_B$ and $\chi_B$ states are essentially
unexplored in QCD~\cite{Heister:2002if}.

\item single-hadron processes such as $\gamma^* \gamma^* \to
\pi^0$, which test the transition from the anomaly-dominated pion
decay constant to the short-distance structure of currents
dictated by the operator-product expansion and perturbative QCD
factorization theorems;

\item hadron pair production processes such as $\gamma^* \gamma
\to \pi^+ \pi^-, K^+ K^-, p \bar p$, which at fixed invariant pair
mass measures the $s \to t$ crossing of the virtual Compton
amplitude \cite{Brodsky:1981rp,BrodskyLepage}.  When one photon is highly virtual,
these exclusive hadron production channels are dual to the photon
structure function $F^\gamma_2(x,Q^2)$ in the endpoint $x \to 1$
region at fixed invariant pair mass.  The leading twist-amplitude
for $\gamma^* \gamma \to \pi^+ \pi^-$ is sensitive to the $1/x -
1/(1-x)$ moment of the $q \bar q$ distribution amplitude
$\Phi_{\pi^+ \pi^-}(x,Q^2)$ of the two-pion system
\cite{Muller:1994fv,Diehl:2000uv}, the time-like extension of
skewed parton distributions.  In addition one can measure the pion
charge asymmetry in $e^+ e^- \to \pi^+ \pi^- e^+ e^-$ arising from
the interference of the $\gamma \gamma \to \pi^+ \pi^-$ Compton
amplitude with the time-like pion form factor
\cite{Brodsky:1971ud}.  At the unphysical point $s =q^2_1= q^2_2 =
0$, the amplitude is fixed by the low energy theorem to the hadron
charge squared.  The ratio of the measured $\gamma \gamma \to
\Lambda \bar \Lambda$ and $\gamma \gamma \to p \bar p$ cross
sections is anomalous at threshold, a fact which may be associated
with the soliton structure of baryons in
QCD~\cite{Sommermann:1992yh,Karliner:2002nk};

\item Exclusive or semi-inclusive channels can also be studied by
coalescence of the produced quarks.  An interesting example is
higher generation final state such as $\gamma \gamma \to B_c \bar
B_c, $ which can have very complex angular
structure~\cite{Brodsky:1985cr}.

\item As pointed out by Hwang~\cite{Hwang}, one can study deeply
virtual Compton scattering on a photon target in $e \gamma$
collisions to determine the light-cone wavefunctions and other
features of the photon~\cite{Brodsky:2000xy}.

\end{enumerate}

\section{Exclusive Two-Photon Annihilation into Hadron
Pairs}

At large momentum transfer, the angular distribution of hadron
pairs produced by photon-photon annihilation are among the best
determinants of the shape of the meson and baryon distribution
amplitudes $\phi_M(x,Q)$ and $\phi_B(x_i,Q),$ which control almost
all exclusive processes involving a hard scale $Q$. The
determination of the shape and normalization of the distribution
amplitudes, which are gauge-invariant and process-independent
measures of the valence wavefunctions of the hadrons, has become
particularly important in view of their importance in the analysis
of exclusive semi-leptonic and two-body hadronic $B$-decays
\cite{BHS,Sz,Beneke:1999br,Keum:2000ph,Keum:2000wi}.  There has
also been considerable progress both in calculating hadron
wavefunctions from first principles in QCD and in measuring them
using diffractive di-jet dissociation.

Hadron pair production from two-photon annihilation
plays a crucial role in unravelling the perturbative and
non-perturbative structure of QCD, first by testing the validity
and empirical applicability of leading-twist factorization
theorems, second by verifying the structure of the underlying
perturbative QCD subprocesses, and third, through measurements of
angular distributions and ratios which are sensitive to the shape
of the distribution amplitudes.  In effect, photon-photon
collisions provide a microscope for testing fundamental scaling
laws of PQCD and for measuring distribution amplitudes.

Two-photon reactions, $\gamma \gamma \to H \bar H$ at large $s =
(k_1 + k_2)^2$ and fixed $\theta_\mathrm{cm}$, provide a
particularly important laboratory for testing QCD since these
cross-channel ``Compton'' processes are the simplest calculable
large-angle exclusive hadronic scattering reactions. The helicity
structure, and often even the absolute normalization can be
rigorously computed for each two-photon channel
\cite{Brodsky:1981rp}. In the case of meson pairs, dimensional
counting predicts that for large $s$, $s^4 d\sigma/dt(\gamma
\gamma \to M \bar M)$ scales at fixed $t/s$ or
$\theta_\mathrm{cm}$ up to factors of $\ln s/\Lambda^2$.  The
angular dependence of the $\gamma \gamma \to H \bar H$ amplitudes
can be used to determine the shape of the process-independent
distribution amplitudes, $\phi_H(x,Q)$.  An important feature of
the $\gamma \gamma \to M \bar M$ amplitude for meson pairs is that
the contributions of Landshoff pitch singularities are power-law
suppressed at the Born level---even before taking into account
Sudakov form factor suppression.  There are also no anomalous
contributions from the $x \to 1$ endpoint integration region.
Thus, as in the calculation of the meson form factors, each
fixed-angle helicity amplitude can be written to leading order in
$1/Q$ in the factorized form $[Q^2 = p_T^2 = tu/s$; $\widetilde
Q_x = \min(xQ,(l-x)Q)]$:
\begin{eqnarray}
\mathcal{M}_{\gamma \gamma\to M \bar M} &= &\int^1_0\, dx \int^1_0
\, dy \nonumber\\
&&\hspace{-2pc}\phi_{\bar M}(y,\widetilde Q_y)
T_H(x,y,s,\theta_\mathrm{cm} \phi_{M}(x,\widetilde Q_x) ,
\end{eqnarray}
where $T_H$ is the hard-scattering amplitude $\gamma \gamma \to (q
\bar q) (q \bar q)$ for the production of the valence quarks
collinear with each meson, and $\phi_M(x,\widetilde Q)$ is the
amplitude for finding the valence $q$ and $\bar q$ with light-cone
fractions of the meson's momentum, integrated over transverse
momenta $k_\perp < \widetilde Q$.  The contribution of non-valence
Fock states are power-law suppressed.  Furthermore, the
helicity-selection rules \cite{Brodsky:1981kj} of perturbative QCD
predict that vector mesons are produced with opposite helicities
to leading order in $1/Q$ and all orders in $\alpha_s$. The
dependence in $x$ and $y$ of several terms in $T_{\lambda,
\lambda'}$ is quite similar to that appearing in the meson's
electromagnetic form factor.  Thus much of the dependence on
$\phi_M(x,Q)$ can be eliminated by expressing it in terms of the
meson form factor.  In fact, the ratio of the $\gamma \gamma \to
\pi^+ \pi^-$ and $e^+ e^- \to \mu^+ \mu^-$ amplitudes at large $s$
and fixed $\theta_{CM}$ is nearly insensitive to the running
coupling and the shape of the pion distribution amplitude:
\begin{equation}{{d\sigma \over dt
}(\gamma \gamma \to \pi^+ \pi^-) \over {d\sigma \over dt }(\gamma
\gamma \to \mu^+ \mu^-)} \sim {4 \vert F_\pi(s) \vert^2 \over 1 -
\cos^2 \theta_\mathrm{cm} }.
\end{equation}
\begin{figure*}[htb]
\begin{center}
\psfig{file=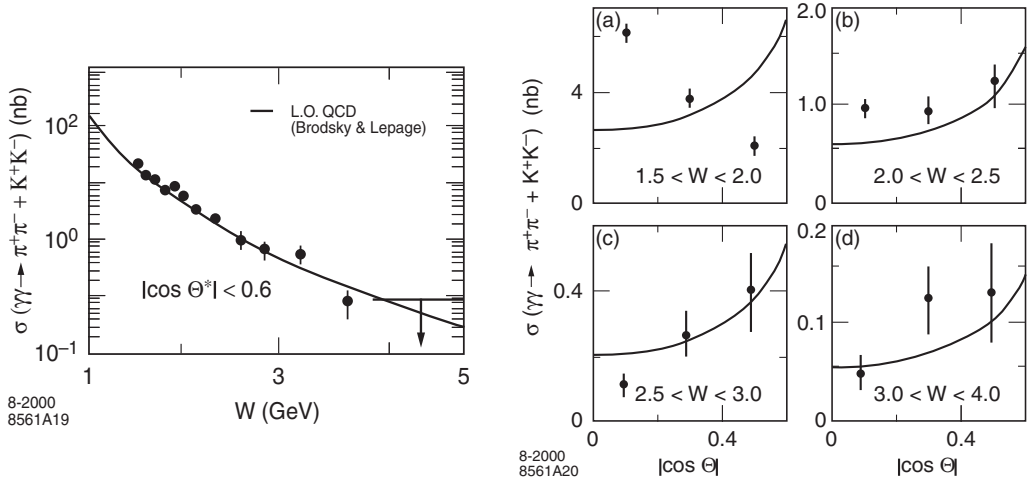,width =.9\textwidth}
\end{center}
\caption[*]{Comparison of the sum of $\gamma \gamma \rightarrow
\pi^+ \pi^-$ and $\gamma \gamma \rightarrow K^+ K^-$ meson pair
production cross sections with the scaling and angular
distribution of the perturbative QCD prediction
\cite{Brodsky:1981rp}.  The data are from the CLEO collaboration
\cite{Dominick:1994bw}. \label{Fig:CLEO}}
\end{figure*}
The comparison of the PQCD prediction for the sum of $\pi^+ \pi^-$
plus $K^+ K^-$ channels with CLEO data \cite{Dominick:1994bw} is
shown in Fig. \ref{Fig:CLEO}.  The CLEO data for charged pion and
kaon pairs show a clear transition to the scaling and angular
distribution predicted by PQCD \cite{Brodsky:1981rp} for $W =
\sqrt(s_{\gamma \gamma} > 2$ GeV. It is clearly important to
measure the magnitude and angular dependence of the two-photon
production of neutral pions and $\rho^+ \rho^-$ in view of the
strong sensitivity of these channels to the shape of meson
distribution amplitudes (see Figs. \ref{Fig:piangle} and
\ref{Fig:rhoangle}).  QCD also predicts that the production cross
section for charged $\rho$-pairs (with any helicity) is much
larger than for that of neutral $\rho$ pairs, particularly at
large $\theta_\mathrm{cm}$ angles.  Similar predictions are
possible for other helicity-zero mesons.  For an alternative model
based on the QCD ``handbag" diagram, see the
references~\cite{Diehl:2001fv}.

The leading-twist QCD predictions for exclusive two-photon
processes such as the photon-to-pion transition form factor and
$\gamma \gamma \to $ hadron pairs are based on rigorous
factorization theorems.  The data from the CLEO collaboration on
$F_{\gamma \pi}(Q^2)$ and the sum of $\gamma \gamma \to \pi^+
\pi^-$ and $\gamma \gamma \to K^+ K^-$ channels are in excellent
agreement with the QCD predictions. It is particularly compelling
to see a transition in angular dependence between the low energy
chiral and PQCD regimes. The problem of setting the
renormalization scale of the coupling for exclusive amplitudes is
discussed in the references~\cite{Brodsky:1997dh}.   The success
of leading-twist perturbative QCD scaling for exclusive processes
at presently experimentally accessible momentum transfer can be
understood if the effective coupling $\alpha_s(Q^*)$ is
approximately constant at the relatively small scales $Q^*$
relevant to the hard scattering amplitudes \cite{Brodsky:1997dh}.
Evidence that the QCD coupling has an IR Fixed point at small
scales has been presented in the references~\cite{Brodsky:2002nb};
The evolution of the quark distribution amplitudes in the
low-$Q^*$ domain also needs to be minimal. Sudakov suppression of
the endpoint contributions is also strengthened if the coupling is
frozen because of the exponentiation of a double logarithmic
series. Conformal symmetry can be used as a guide to organize the
hard scattering calculation and to determine the leading
contributions to the hadron distribution
amplitudes~\cite{Brodsky:1980ny,Brodsky:2000cr,Braun:1999te,Belitsky:2002kj,%
Brodsky:2003px}.

\begin{figure}[htbp]
\begin{center}
\psfig{file=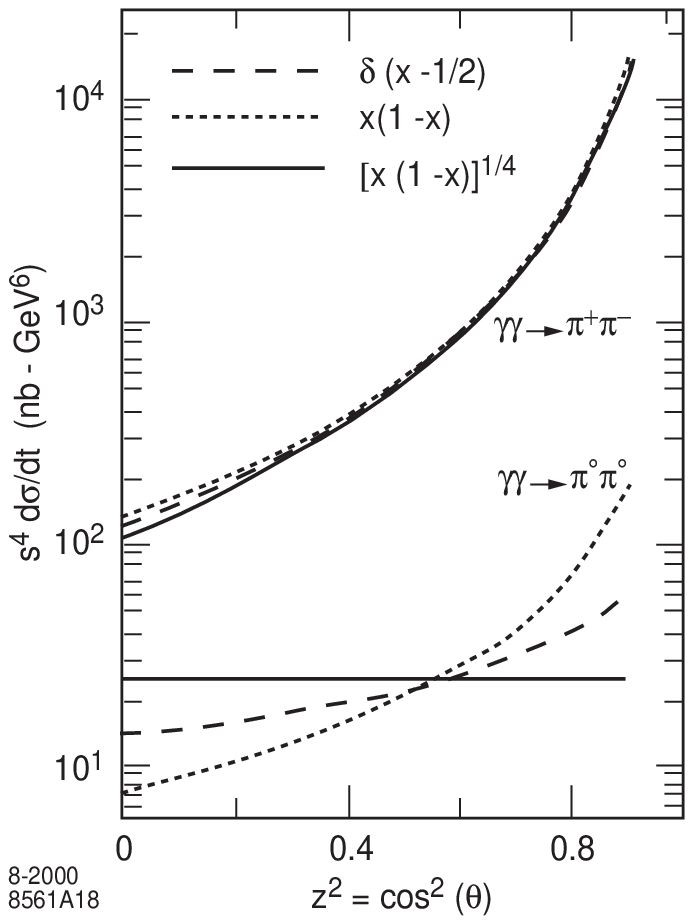,width=.48\columnwidth}
\end{center}
\caption[*]{ Predictions for the angular distribution of the
$\gamma\gamma\rightarrow \pi^+\pi^-$ and $\gamma \gamma
\rightarrow \pi^0 \pi^0$ pair production cross sections for three
different pion distribution amplitudes \cite{Brodsky:1981rp}.
\label{Fig:piangle}}
\bigskip
\begin{center}
\psfig{file=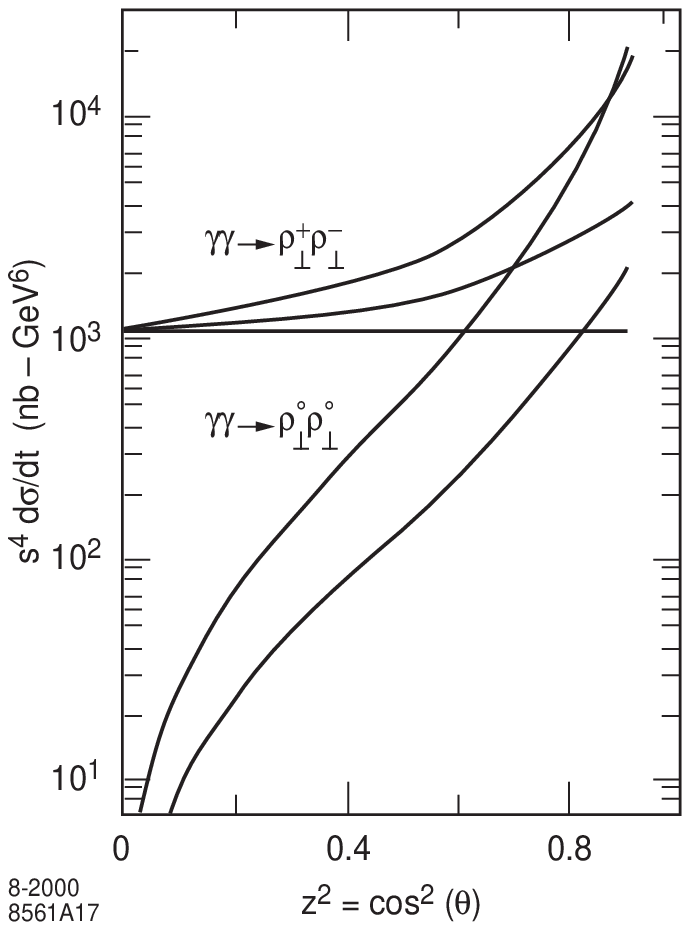,width=.48\columnwidth}
\end{center}
\caption[*]{Predictions for the angular distribution of the
$\gamma\gamma\rightarrow \rho^+\rho^-$ and $\gamma \gamma
\rightarrow \rho^0 \rho^0$ pair production cross sections for
three different $\rho$ distribution amplitudes as in Fig.
\ref{Fig:piangle} \cite{Brodsky:1981rp}. \label{Fig:rhoangle}}
\end{figure}

\begin{figure}[!t]
\begin{center}
\psfig{file=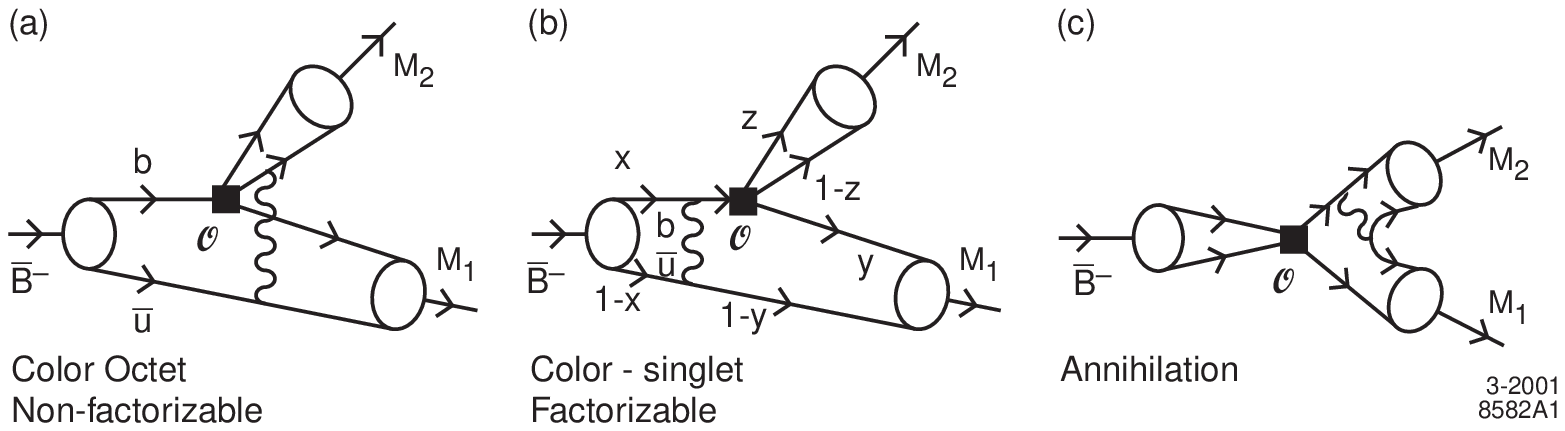,width=5in}
\end{center}
\caption[*]{Three representative contributions to exclusive $B$
decays to meson pairs in PQCD.  The operators $\cal O$ represent
the QCD-improved effective weak interaction. \label{fig:B}}
\end{figure}

As noted above, the analysis of exclusive $B$ decays has much in common with the
analysis of exclusive two-photon reactions~\cite{Brodsky:2001jw}. For example,
consider the three representative contributions to the decay of a $B$ meson to
meson pairs illustrated in Fig. \ref{fig:B}.  In Fig. \ref{fig:B}(a) the weak
interaction effective operator $\mathcal{O}$ produces a $ q \bar q$ in a color
octet state.  A gluon with virtuality $Q^2 = \mathcal{O} (M_B^2)$ is needed to
equilibrate the large momentum fraction carried by the $b$ quark in the $\bar B$
wavefunction. The amplitude then factors into a hard QCD/electroweak subprocess
amplitude for quarks which are collinear with their respective hadrons: $T_H([b(x)
\bar u(1-x)] \to [q(y) \bar u(1-y)]_1 [q(z) \bar q(1-z)]_2)$ convoluted with the
distribution amplitudes $\phi(x,Q)$ \cite{BrodskyLepage} of the incident and final
hadrons:
\begin{eqnarray*}
\mathcal{M}_\mathrm{octet}(B \to M_1 M_2) &= &\int^1_0\, dz
\int^1_0\, dy \int^1_0\,
dx\\
&&\hspace{-3pc}\phi_B(x,Q) T_H(x,y,z) \phi_{M_1}(y,Q)
\phi_{M_2}(z,Q).
\end{eqnarray*}
Here $x = k^{+}/p^{+}_H = (k^0+ k^z) /(p^0_H + p^z_H)$ are the
light-cone momentum fractions carried by the valence quarks.

Baryon pair production in two-photon annihilation is also an important testing
ground for QCD~\cite{Brodsky:1981rp,Lepage:1980fj}.  The calculation of $T_H$ for
Compton scattering requires the evaluation of 368 helicity-conserving tree
diagrams which contribute to $\gamma (qqq) \to \gamma^\prime (qqq)^\prime$ at the
Born level and a careful integration over singular intermediate energy
denominators \cite{Farrar:1990qj,Kronfeld:1991kp,Guichon:1998xv}.  Brooks and
Dixon \cite{Brooks:2000nb} have completed a recalculation of the proton Compton
process at leading order in PQCD, extending and correcting earlier work.  It is
useful to consider the ratio $ d\sigma/dt(\gamma \gamma \to \bar p p)/
d\sigma/dt(e^+ e^- \to \bar p p)$ since the power-law fall-off, the normalization
of the valence wavefunctions, and much of the uncertainty from the scale of the
QCD coupling cancel.  The scaling and angular dependence of this ratio is
sensitive to the shape of the proton distribution amplitudes.  The perturbative
QCD predictions for the phase of the Compton amplitude phase can be tested in
virtual Compton scattering by interference with Bethe-Heitler processes
\cite{Brodsky:1972vv}.

The cross section for $\gamma \gamma \to p \bar p$ has been measured at LEP by the
OPAL and L3 collaborations~\cite{Abbiendi:2002bx,Achard:2003jc,Braccini:2003rm}.
The results are in essential agreement with perturbative QCD scaling and the
predictions for the angular dependence.  However, the normalization of the PQCD
three-quark predictions~\cite{Farrar:gv,Millers:ca} appear to underestimate the
measured exclusive cross section by an order of magnitude.  This could be evidence
for an effective quark-diquark structure of the proton
wavefunction~\cite{Diehl:2002yh,Berger:2003ve}. However, the normalization
discrepancy with the leading-twist QCD predictions could be due to several
reasons. (1) Note that the rate is proportional to $\alpha^4_s.$ The QCD coupling
is assumed to have a nominal value $\alpha_s \sim 0.27,$~\cite{Millers:ca} which
could be an underestimate of the effective value of the running QCD coupling at
the relevant scales. (2) The leading order perturbative QCD predictions lead to
real amplitudes. The imaginary parts which appear from the analytic continuation
of the QCD coupling and final state QCD interactions can also give a significant
correction, as is known to be the case in $e^= e^- \to p \bar
p$~\cite{Brodsky:2003gs}. (3) The relatively slow fall-off of the Pauli form
factor as observed in polarization transfer experiments at Jefferson lab, suggests
significant higher twist contributions.   All of these arguments suggest that the
most reliable test of QCD are obtained from the ratio: $ d\sigma/dt(\gamma \gamma
\to \bar p p)/ d\sigma/dt(e^+ e^- \to \bar p p)$.  In addition measurements of the
proton polarization single-spin asymmetry provides important information on the
phase of the timelike pair production amplitudes~\cite{Brodsky:2003gs}.

It is also interesting to measure baryon and isobar pair
production in two photon reactions near threshold.  Ratios such as
$\sigma(\gamma \gamma \to \Delta^{++} \Delta^{--})/\sigma(\gamma
\gamma \to \Delta^{+} \Delta^{-})$ can be as large as $16:1$ in
the quark model since the three-quark wavefunction of the $\Delta$
is expected to be symmetric.  Such large ratios would not be
expected in soliton models \cite{Sommermann:1992yh} in which
intermediate multi-pion channels play a major role.

Pobylitsa {\em et al.} \cite{Pobylitsa:2001cz} have shown how the
predictions of perturbative QCD can be extended to processes such
as $\gamma \gamma \to p \bar p \pi$ where the pion is produced at
low velocities relative to that of the $p$ or $\bar p$ by
utilizing soft pion theorems in analogy to soft photon theorems in
QED.  The distribution amplitude of the $p \pi$ composite is
obtained from the proton distribution amplitude from a chiral
rotation.  A test of this procedure in inelastic electron
scattering at large momentum transfer $e p \to p \pi$ and small
invariant $p^\prime \pi$ mass has been remarkably successful. Many
tests of the soft meson procedure are possible in multiparticle
$e^+ e^-$ and $\gamma \gamma$ final states.

One of the formidable challenges in QCD is the calculation of
non-perturbative wavefunctions of hadrons from first principles.
The calculations of the pion distribution amplitude by Dalley
\cite{Dalley:2000dh} and by Burkardt and Seal
\cite{Burkardt:2001mf} using light-cone and transverse lattice
methods is particularly encouraging.  The predicted form of
$\phi_\pi(x,Q)$ is somewhat broader than but not inconsistent with
the asymptotic form favored by the measured normalization of $Q^2
F_{\gamma \pi^0}(Q^2)$ and the pion wavefunction inferred from
diffractive di-jet production.

Clearly much more experimental input on hadron wavefunctions is
needed, particularly from measurements of two-photon exclusive
reactions into meson and baryon pairs at the high luminosity $B$
factories.  For example, as shown in Fig. \ref{Fig:piangle}, the
ratio
\begin{displaymath}
{{d\sigma \over dt }(\gamma \gamma \to \pi^0 \pi^0) / {d\sigma
\over dt}(\gamma \gamma \to \pi^+ \pi^-)}
\end{displaymath}
is particularly sensitive to the shape of pion distribution
amplitude.  At fixed pair mass, and high photon virtuality, one
can study the distribution amplitude of multi-hadron states
\cite{Diehl:2000uv}.  Two-photon annihilation will provide much
information on fundamental QCD processes such as deeply virtual
Compton scattering and large angle Compton scattering in the
crossed channel.  I have also emphasized the interrelation between
the wavefunctions measured in two-photon collisions and the
wavefunctions needed to study exclusive $B$ and $D$ decays.

Much of the most interesting two-photon annihilation physics is
accessible at low energy,  high luminosity $e^+ e^-$ colliders,
including measurements of channels important in the light-by-light
contribution to the muon $g$--2 and the study of the transition
between threshold production controlled by low-energy effective
chiral theories and the domain where leading-twist perturbative
QCD becomes applicable.

The threshold regime of hadron production in photon-photon and
$e^+ e^-$ annihilation, where hadrons are formed at small relative
velocity, is particularly interesting as a test of low energy
theorems, soliton models, and new types of resonance production.
Such studies will be particularly valuable in double-tagged
reactions where polarization correlations, as well as the photon
virtuality dependence, can be studied.

\section{New Calculational Methods}

The light-front quantization of gauge theories can be carried out
in an elegant way using the Dirac method to impose the light-cone
gauge constraint and eliminate dependent degrees of
freedom~\cite{Srivastava:2000cf}.  Unlike the case in equal-time
quantization, the vacuum remains trivial.  Since only physical
degrees of freedom appear, unitarity is maintained.  One can
verify the QCD Ward identities for the physical light-cone gauge
and compute the QCD $\beta$ function. Srivastava and
I~\cite{Srivastava:2002mw} have extended the light-front
quantization procedure to the Standard Model. The spontaneous
symmetry breaking of the gauge symmetry is due to a zero mode of
the scalar field rather than vacuum breaking.  The Goldstone
component of the scalar field provides mass to the $W^\pm$ and
$Z^0$ gauge bosons as well as completing its longitudinal
polarization.  The resulting theory is free of Faddeev-Popov
ghosts and is unitary and renormalizable.  The resulting rules
give an elegant new way to compute Standard model processes using
light-front Hamiltonian theory.

The light-front method suggests the possibility of developing an
``event amplitude generator" for high energy processes such as
photon-photon collisions by calculating amplitudes for specific
parton spins using light-front time-ordered perturbation
theory~\cite{Brodsky:2001ww}.  The positivity of the $k^+$
light-front momenta greatly constrains the number of contributing
light-front time orderings. Since particle states are labelled by
the spin projection $S_z$, total angular momentum constraints are readily
implemented. The renormalized
amplitude can be obtained diagram by diagram by using the ``alternating
denominator" method which automatically subtracts the relevant
counterterm.  The DLCQ method also provides a simple way to
discretized the light-front momentum variables, while maintaining
frame-independence.  The resulting renormalized amplitude can be
convoluted with the light-front wavefunctions to simulate
hadronization and hadron matrix elements.

\section*{Acknowledgments}
I thank Professor Clem Heusch for organizing this workshop.  Work
supported by the Department of Energy under contract number
DE-AC03-76SF00515.

\end{document}